\documentclass{article}
\usepackage{color,soul}
\usepackage{graphicx}
\usepackage{makecell}
\usepackage{amsmath}
\usepackage{epsf,hyperref}
\usepackage{amssymb,ComplexSystems}
\usepackage{booktabs}
\usepackage{subcaption}
\usepackage{xcolor}
\begin{document}

\title{Transfer Learning for Node Regression Applied to Spreading Prediction} 

\author{\authname{Sebastian Me\v{z}nar}\\[2pt] 
\authadd{Jo\v{z}ef Stefan Institute, Jamova 39, Ljubljana, Slovenia}\\
\and
\authname{Nada Lavra\v{c}}\\ [2pt]
\authadd{Jo\v{z}ef Stefan Institute, Jamova 39, Ljubljana, Slovenia}\\
\authadd{University of Nova Gorica, Glavni trg 8, Vipava, Slovenia} \\
\and
\authname{Bla\v{z} \v{S}krlj}\\[2pt]
\authadd{Jo\v{z}ef Stefan Institute, and Jo\v{z}ef Stefan International Postgraduate School}\\
\authadd{Jamova 39, Ljubljana, Slovenia}\\}

\markboth{Complex Systems}{Transfer Learning for Node Regression Applied to Spreading Prediction}

\maketitle

\begin{abstract}
Understanding how information propagates in real-life complex networks yields a better understanding of dynamic processes such as misinformation or epidemic spreading. The recently introduced branch of machine learning methods for learning node representations offers many novel applications, one of them being the task of spreading prediction addressed in this paper. We explore the utility of the state-of-the-art node representation learners when used to assess the effects of spreading from a given node, estimated via extensive simulations. Further, as many real-life networks are topologically similar, we systematically investigate whether the learned models generalize to previously unseen networks, showing that in some cases very good model transfer can be obtained. This work is one of the first to explore transferability of the learned representations for the task of node regression; we show there exist pairs of networks with similar structure between which the trained models can be transferred (zero-shot), and demonstrate their competitive performance. To our knowledge, this is one of the first attempts to evaluate the utility of zero-shot transfer for the task of node regression.
\end{abstract}

\begin{keywords}
epidemics; neural networks; machine learning; spreading; transfer learning
\end{keywords}

\section{Introduction}\label{sec:intro}
Spreading of information or disease spreading are examples of common phenomena of spreading. Modeling the spreading process and spreading prediction has many practical and potentially life-saving applications, including the creation of better strategies for stopping the spreading of misinformation on social media or stopping an epidemic. Further, companies can analyze spreading to create better strategies for marketing their product~\cite{Guille2013diffusion,nowzari2016analysis}. Spreading analysis can also be suitable for analysis of e.g., fire spreading, implying large practical value in terms of insurance cost analysis~\cite{kacem2017small}. Analysis of spreading is commonly studied via extensive simulations~\cite{kesarev2018parallel}, exploiting ideas from statistical mechanics to better understand both the extent of spreading, as well as its speed~\cite{dong2018studies}. 

While offering high utility, reliable simulations of spreading processes can be \emph{expensive} when performed on larger networks. This issue can be addressed by employing \emph{machine learning}-based modeling techniques~\cite{wu2020comprehensive}. The contributions of this work are multi-fold and can be summarized as follows.
\begin{enumerate}
    \item We propose an efficient network node regression algorithm named CaBoost, which achieves state-of-the-art performance for the task of spreading prediction against strong baselines such as graph neural networks.
    \item We demonstrate that machine learning-based spreading prediction can be utilized for fast screening of potentially problematic nodes, indicating that this branch of methods is complementary to the widely adopted simulation-based approaches.
    \item We investigate to what extent the models, learned on a given network A are \emph{transferable} to a network B, and what type of structural features preserve this property the best. This hypothesis assumes that structure-only properties could be sufficient for model transfer in some cases. We demonstrate that transfer learning for node regression is possible, albeit only across topologically similar networks.
\end{enumerate}

This work extends the paper `Prediction of the effects of epidemic spreading with graph neural networks'~\cite{meznar2020spreading} from the Complex Networks 2020 conference by testing more approaches, using more simulation data, and using different node centralities. Additionally, this work tests if centrality-based features can be used for \emph{zero-shot transfer learning}. 

The remainder of this work is structured as follows. Section~\ref{sec:related} presents the related work which led to the proposed approach. In Section~\ref{sec:proposed-method} we present the proposed methodology, where we re-formulate the task, present centrality data used to create features for our learners and show how we approached transfer learning. In Section~\ref{sec:empirical-eval} we present the datasets, the experimental setting, interpretation of the predictions, and the results of the empirical evaluation and transfer learning. We conclude the paper with the discussion in Section~\ref{sec:discussion-conclusion}.

\section{Related work}~\label{sec:related}
This section presents the relevant related work. It starts by discussing the notion of contagion processes, followed by an overview of graph neural networks and transfer learning.

\subsection{Analysis of spreading processes}
The study of spreading processes on networks is a lively research endeavor~\cite{nowzari2016analysis}. Broadly, spreading processes can be split into two main branches, namely, the simulation of \emph{epidemics} and \emph{opinion dynamics}. The \emph{epidemic spreading} models can be classical or network-based. The classical models are, for example, systems of differential equations that do not account for a given network's topology. Throughout this work, we are interested in extensions of such models to real-life network settings.

One of the most popular spreading models extended to networks is the Susceptible - Infected - Recovered (SIR) model~\cite{Kermack1927Epidemics}. The spread of the pandemic in the SIR model is dictated by parameters $\beta$ known as the infection rate and $\gamma$ known as the recovery rate. Nodes in this model can have one of three states (Susceptible, Infected, Recovered). 
SIR assumes that if an infected node comes into contact with a susceptible one during a generic iteration, the susceptible node becomes infected with probability $\beta$. In each iteration after getting infected, a node can recover with probability $\gamma$ (only transitions from S to I and from I to R are allowed). 

Other related models include, for example, SEIR, SEIS, SWIR\footnote{Where S-Susceptible, I-Infected, R-Recovered, E-Exposed, and W-Weakened.}. Further, one can also study the role of cascades~\cite{Kempe2003cascades} or the Threshold model~\cite{Granovetter1979threshold}. For the interested reader, multiple other approaches are summarised in ~\cite{li2017survey}.

\subsection{Machine learning on networks}
\label{sec:learning-graphs}
Learning by propagating information throughout a given network has already been considered by the approaches such as label propagation~\cite{zhu2002learning}. However, in the recent years, approaches that jointly exploit both the adjacency structure of a given network alongside features assigned to its nodes are becoming prevalent in the network learning community. The so-called graph neural networks have re-surfaced with the introduction of the Graph Convolutional Networks (GCN)~\cite{kipf2016semi}; an idea where the normalized adjacency matrix is directly multiplied with the feature space and effectively represents a neural network layer. Multiple such layers can be stacked to obtain better approximation power/performance. One of the most recent methods from this branch are the Graph Attention Networks (GAT)~\cite{velickovic2018graph}, an extension of GCNs, extended with the idea of neural \emph{attention}. Here, part of the neural network focuses on particular parts of the adjacency space, offering more robust and potentially better performance. 

Albeit being in widespread use, graph neural networks are not necessarily the most suitable choice when learning solely from the network adjacency structure. For such tasks, methods such as node2vec~\cite{grover2016node2vec}, SGE~\cite{sge}, SNoRe~\cite{meznar2020snore}, and DeepWalk~\cite{Perozzi2014deepwalk} were developed. This branch of methods corresponds to what we refer to as \emph{structural} representation learning. In our work, we focus mostly on learning this type of representations using network centrality information.

Note that although graph neural networks are becoming the prevailing methodology for learning from \emph{feature-rich} complex networks, it is not clear whether they perform competitively to the more established structural methods if the feature space is derived solely from a given \emph{network's structure}.

\subsection{Transfer learning}
The main bottleneck of spreading effect prediction are the expensive simulations. While the number of simulations can be reduced by using machine learning, computation of a large fraction of them might still be infeasible on larger networks. One of the solutions for this problem is \textbf{transfer learning}~\cite{zhuang2020comprehensive}. Transfer learning can be performed in at least two main ways, by fine-tuning a pretrained model (few-shot learning) or by using a model trained on a related problem (zero-shot learning). In our work, we focus on \emph{zero-shot learning}.

In the recent years, new approaches were proposed for transfer learning on networks. These adopted mostly fine-tune pretrained graph neural networks to solve the proposed problem. An example of this is prediction of optoelectronic properties of conjugated oligomers~\cite{lee2021transfer}, where the graph neural network is trained on short oligomers and then fine-tuned by using 100 long ones. The results showed that the fine-tuned model performed much better and needed only a small sample of extra data to improve performance by a margin of 37\%. Another approach tries to predict traffic congestion of a network with a small amount of historical data by training a recurrent neural network on a traffic network with a lot of historic data~\cite{Mallick2020TransferLW}. Zero-shot learning is less popular in the network setting, but few approaches that exist closely follow the paradigms of zero-shot learning~\cite{wang2019zerosurvey}. One such example ~\cite{Kampffmeyer_2019_CVPR} proposes a Dense Graph Propagation module that adds direct links between distant nodes in the knowledge networks to exploit their hierarchical structure. Finally, pre-training graph neural networks is becoming an active research venue, demonstrating that this problem is possible to solve via systematic selection of pre-training data~\cite{qiu2020gcc,hu2019strategies}.
 
When transferring knowledge between different networks, one must carefully craft features that are independent of a single network and represent global node characteristics. Because of this, embedding methods such as node2vec~\cite{grover2016node2vec} and SNoRe~\cite{meznar2020snore} cannot be used since they learn node representation through node indices encountered during random walks within the input network. On the other hand, transfer learning with graph neural networks is an active field of research with most approaches fine-tuning the pre-trained models.

\section{Proposed methodology}\label{sec:proposed-method}
After introducing the task of spreading prediction, a brief methodology overview is presented. After this we present the creation of target variables, extraction of node features from the network, training of machine learning models, and the transfer of models between the networks. 

\subsection{Task formulation}\label{sec:task-def}
In the case of a pandemic, the intensity of disease spreading can be summarized with the following three values: the maximum number of infected people (the peak), the time it takes to reach the maximum number of infected people, and the total number of infected people. Let us consider why these three values are important. Knowing the maximum number of infected people during a pandemic will help us to better prepare for the crisis, as it provides a good estimation of how many resources (e.g., hospital beds) will need to be allocated to patients. In another example domain, companies might want to create marketing campaigns on platforms such as Twitter and target specific users to reach a certain number of retweets that are needed to become trending. The time needed to reach the peak is important e.g., to estimate the high time for developing a cure for the disease, or e.g., to estimate the high time for stopping the spread of misinformation on social media. Finally, the total number of infected nodes during an epidemic is important for assessing the damage, or in another example, for estimating how many computers were infected by some malware.

In our work, we focus on predicting the maximum number of infected nodes and the time needed to reach this number. We create target data by simulating epidemics from each node with the \emph{SIR diffusion model}~\cite{ndlib} and identify the number of nodes, as well as the time when the maximum number of nodes are infected. We aggregate the generated target data by taking the mean values for each node (expected time and infection numbers). Finally, we normalize the data as follows. We divide the maximum number of infected nodes by the number of nodes in the observed network. This normalization is suitable since the maximum number of infected nodes can not exceed the number of nodes in the network. 
This normalization intuitively offers insight into the \emph{percentage} of infected nodes, regardless of scale.
The upper bound for the time when the maximum number of nodes are infected does not exist. Because of this, we divide the time when the maximum number of nodes are infected with the maximum from the observed data. In practice we might not have this maximum, so a suitable number must be chosen by e.g., examining prior events. Furthermore, in practice, we create the target data as described above and save the number by which we divide (normalization constant). After prediction, we multiply the prediction with this number to find the `real-world' equivalent. 

\subsection{Methodology overview}
\begin{figure}[b!]
  \centering
  \includegraphics[width=\linewidth]{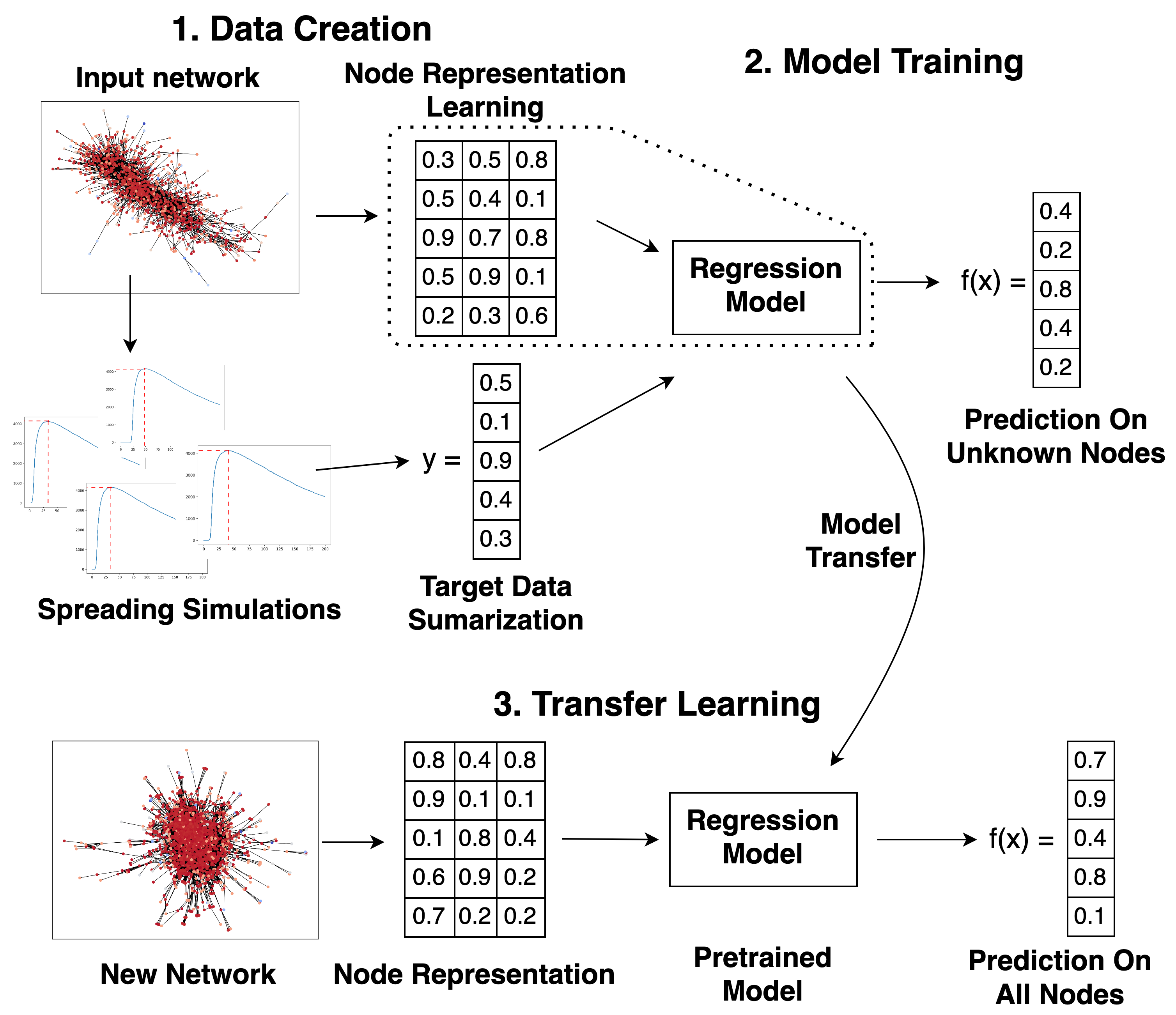}
  \caption{Overview of the proposed methodology.}
  \label{fig:overview}
\end{figure}

The overview of the proposed methodology is presented in Figure~\ref{fig:overview}. The methodology is composed of three main steps: input data creation, model training, and transfer learning. 

In the first step, we create target variables and node features from the starting (initial) network. Target variables are created by running simulations on the starting network followed by their summarization and processing. While target variable creation can be performed in the same way on each network, feature creation depends on the learner. Embeddings and feature extraction methods generate features stored in tabular format, which is then used for model training. Graph neural networks are on the other hand end-to-end learners, meaning they learn features and perform regression at the same time.

In the second step, we train a regression model using the extracted features and the target variables we created. We use this model to predict the target variables for the \emph{unknown nodes}. 

The model learned in the second step is used in the last step to predict target variables of nodes from a new (unknown - unobserved) network. For such predictions, we first extract features from the new network and then use them on the pretrained model. These predictions transfer knowledge from the first network to the nodes of the second one.

\subsection{Training data creation}\label{sec:training-data}
The first part of the methodology addresses input data generation. Intuitively, the first step simulates epidemic spreading from individual nodes of a given network to assess both the time required to reach the maximum number of infected nodes, as well as the maximum number itself. In this work, we leverage the widely used SIR model~\cite{Kermack1927Epidemics}~\cite{ndlib} to simulate epidemics, formulated as follows.
\begin{align*}
    \frac{dS}{dt} &= - \frac{\beta \cdot S \cdot I}{N} \\
    \frac{dI}{dt} &= \frac{\beta \cdot S \cdot I}{N} - \gamma \cdot I \\
    \frac{R}{dt} &= \gamma \cdot I,
\end{align*}
\noindent where S represents the number of susceptible, R the number of recovered and I the number of infected individuals. Spreading is governed by input parameters $\gamma$ and $\beta$. The SIR model is selected due to many existing and optimized implementations that are adapted from systems of differential equations to networks~\cite{Guille2013diffusion}. We use NDlib~\cite{ndlib} to simulate epidemics based on the \emph{SIR diffusion model}.

Target data creation results in two real values for each node. We attempt to \emph{predict} these two values. The rationale for the construction of such predictive models is that they are potentially much faster than simulating multiple spreading processes for each node (prediction time is the bottleneck) and can give more insight into \emph{why} some nodes are more `dangerous'. 

The predictive task can be formulated as follows. Let $G = (V,E)$ represent the considered network. We are interested in finding the mapping $f: V \rightarrow [0, 1]$ from the set of nodes $V$ to the set of real values that represent e.g., the maximum number of infected individuals if the spreading process is started from a given node $u \in V$. Thus, $f$ corresponds to the process of \textbf{node regression}.

\subsection{Learning on the same network: Prediction with simulation data}\label{sec:machine}
The models we considered can broadly be split into two main categories: graph neural networks and propositional learners. The main difference between the two is that the graph neural network learners, such as GAT~\cite{velickovic2018graph} and GIN~\cite{xu2018powerful} simultaneously exploit both the structure of a network, as well as \emph{node features}, whilst the propositional learners take as input only the constructed feature space (and not the adjacency matrix). As an example, the feature space is passed through the GIN's structure via the update rule that can be stated as:
\begin{equation*}
    \boldsymbol{h}_v^{(k)} = \textsc{MLP}^{(k)}\bigg ( \big ( 1 + \epsilon^{(k)} \big ) \cdot \boldsymbol{h}_v^{(k-1)} +  \sum_{\boldsymbol{u} \in V(v)} \boldsymbol{h}_u^{(k-1)} \bigg ),
\end{equation*}
\noindent where $\textsc{MLP}$ corresponds to a multilayer perceptron, $\epsilon$ a hyperparameter, $\boldsymbol{h}_u^{(k)}$ the node $u$'s representations at layer $k$ and $V(v)$ the $v$-th node's neighbors. We test both graph neural networks and propositional learners as it is to our knowledge not clear whether direct incorporation of the adjacency matrix offers any superior performance, as the graph neural network models are computationally more expensive. The summary of considered learners is presented in Table~\ref{tab:learners}.

\begin{table}[b!]
    \centering
    \caption{Summary of the considered learners with descriptions, where $\boldsymbol{A}$ denotes the adjacency matrix and $\boldsymbol{F}$ the feature matrix.}\vspace{0.2cm}
    \resizebox{\columnwidth}{!}{
        \begin{tabular}{cll}
        \hline
            Input  & Learner & Method description  \\ \hline
            $\boldsymbol{A}, \boldsymbol{F}$ & GAT & Graph Attention Networks \\
            $\boldsymbol{A}, \boldsymbol{F}$ & GIN & Graph Isomorphism Networks \\
            $\boldsymbol{A}$ & node2vec + XGBoost & node2vec-based features + XGBoost \\
            $\boldsymbol{A}, \boldsymbol{F}$ & node2vec + features + XGBoost & node2vec-based features + centrality based features + XGBoost \\
            $\boldsymbol{A}$ & SNoRe + XGBoost & SNoRe-based features + XGBoost \\
            $\boldsymbol{A}, \boldsymbol{F}$ & SNoRe + features + XGBoost & SNoRe-based features + centrality based features + XGBoost \\
            $\boldsymbol{F}$ & CaBoost & XGBoost trained solely on centrality based features \\
            \hline
    \end{tabular}}
    \label{tab:learners}
\end{table}

As the considered complex networks do not possess \emph{node attributes}, we next discuss which features, derived solely from the network structure were used in the considered state-of-the-art implementations of GAT~\cite{velickovic2018graph} and GIN~\cite{xu2018powerful}, or concatenated to an embedding generated using node2vec~\cite{grover2016node2vec} or SNoRe~\cite{meznar2020snore} for use in XGBoost. Further, we also test models where only the constructed structural features are considered, as well as a standalone method capable of learning node representations, combined with the XGBoost~\cite{xgboost} classifier. In this work, we explore whether \emph{centrality-based} descriptions of nodes are suitable for the considered learning task. The rationale for selecting such features is that they are potentially fast to compute and entail global relation of a given node w.r.t. the remaining part of the network. The centralities, computed for each node, are summarized in Table~\ref{tab:centralities}. These centralities are then normalized and concatenated to create features used with some learners. In Section~\ref{sec:results} we refer to the XGBoost model trained with these features as CaBoost, which is one of the contributions of this work.

\begin{table}[t!]
    \centering
        \caption{Summary of the centralities considered in our work. 
        }\vspace{0.2cm}
        \resizebox{1\columnwidth}{!}{
    \begin{tabular}{cll}
    \hline
       Centrality  & \makecell{Time \\complexity}& \makecell{Description} \\ \hline
        Degree centrality~\cite{rodrigues2019network} & \makecell{$\mathcal{O}(|E|)$} & \makecell{The number of edges of a given node.}\\ \hline
        Eigenvector centrality~\cite{rodrigues2019network} & \makecell{$\mathcal{O}(|V|^3)$} & \makecell{Importance of the node, where nodes are\\ more important if they are connected to \\other important nodes. This can be \\ calculated using the eigenvectors of the\\ adjacency matrix.} \\ \hline
        PageRank~\cite{Page1999ThePC} & \makecell{$\mathcal{O}(|E|)$} & \makecell{Probability that a random walker \\ is at a given node.} \\ \hline
        Average Out-degree & \makecell{$\mathcal{O}(|V|\cdot w \cdot\overline{s})$}  & \makecell{The average out-degree of nodes \\ encountered during $w$ random walks \\ of mean length $\overline{s}$.}\\ \hline
        \makecell{Number of second neighbors} & \makecell{$\mathcal{O}(|V|\cdot|E|)$} & \makecell{Number of nodes that are neighbors to\\ neighbors of a given node. This number\\ is between $0$ and $|V|$.} \\ \hline
    \end{tabular}}
    \label{tab:centralities}
\end{table}

During model training, we minimize the mean squared error (MSE) between the prediction $f(u)$ and the observed state $y_u$, which is defined for the $u$-th node as follows:
\begin{equation*}
    MSE =  1/|N|\sum_{u \in N} (f(u)-y_u)^2.
\end{equation*}

In Section~\ref{sec:results} we use the root mean squared errors (RMSE), defined as follows:
\begin{equation*}
    RMSE = \sqrt{MSE},
\end{equation*}
\noindent to present the results.

\subsection{Transfer learning from other networks}
In Section~\ref{sec:machine} we use the centrality data to create the features used for model training. Since these features represent (normalized) global characteristics of nodes and not the specific relations between them (as for example in node2vec or SNoRe), they have the advantage of being transferable between different networks. This gives us the ability to train a model on one network and use it for prediction on a different network.

In this work, we use the approach outlined in the following paragraphs to train and test a regression model for transfer learning. We will use the term \emph{training network} to highlight the network used for training the model, and \emph{test network} as the network composed of nodes used in prediction of spreading effects.

First we create simulations with nodes from the training network as patient(s) zero, and create target variables as shown in Sections~\ref{sec:task-def} and~\ref{sec:training-data}. After this, we create centrality based features and use them to train the CaBoost model from Section~\ref{sec:machine}. To predict target variables of the test network, we generate its centrality based features and use them with the previously trained model.

In Section~\ref{sec:transfer-results} we use the following methodology to benchmark the performance of transfer learning models. First we create target variables and centrality based features of all networks. Then we normalize the features and use all instances to train one CaBoost model for each network. Transfer learning scores are then calculated for each model and each (different) network as the RMSE between the predicted values and target variables. We use five-fold cross-validation as the baseline score for each network.

We represent the performance of transfer learning as a heatmap. The \textbf{columns of the heatmap represent the test networks}, while the \textbf{rows represent the training network} used to create the model. The values on the diagonal represent the RMSE values of the five-fold cross-validation. The other values represent the \textbf{transfer learning score} (score on the test network) divided by the baseline score. The values can be interpreted as the decrease in performance if we use a model trained on another network relative to the estimated performance on the initial network obtained in the process of five-fold cross-validation.

\section{Empirical evaluation}\label{sec:empirical-eval}
In this section, we present the baselines and datasets used for evaluation and show the empirical results of the approaches outlined in Section~\ref{sec:proposed-method}. We also present how predictions from SNoRe+features model can be explained with  SHAP~\cite{lundberg2017shap}.

\subsection{Baselines for regression (initial network)}
\label{sec:baselines}
We compared the results of proposed method to the following five baselines:
\begin{description}
\item[\emph{Random baseline}] creates an embedding of size $|N|\times 64$ with random numbers drawn from $\textrm{Unif}(0,1)$. We use this embedding as the input data for the XGBoost model.
\item[\emph{node2vec}~\cite{grover2016node2vec}] learns a low dimensional representation of nodes that maximizes the likelihood of neighborhood preservation using random walks. During testing, we use the default parameters.
\item[\emph{SNoRe}~\cite{meznar2020snore}] learns an interpretable representation of nodes based on the similarity between their neighborhoods. These neighborhoods are created with short random walks. During testing, we use the default parameters.
\item[\emph{GAT}~\cite{velickovic2018graph}] includes the attention mechanism that helps learn the importance of neighboring nodes. In our tests, we use the implementation from PyTorch Geometric~\cite{Fey2019ptg}.
\item[\emph{GIN}~\cite{xu2018powerful}] learns a representation that can provably achieve the maximum discriminative power. In our tests, we use the implementation from PyTorch Geometric~\cite{Fey2019ptg}.
\end{description}

\noindent For comparison we also add the averaged simulation error. We calculate this error with the RMSE formula, where $y=0$ and $f(u)$ is the mean absolute difference between simulation results and their mean value. This baseline corresponds to the situation, where only a single simulation would be used to approximate the expected value of multiple ones (the goal of this work).

\subsection{Experimental setting}
\label{sec:setting}
\begin{figure}[t!]
  \begin{subfigure}{0.49\textwidth}
    \centering
    \includegraphics[width=\textwidth]{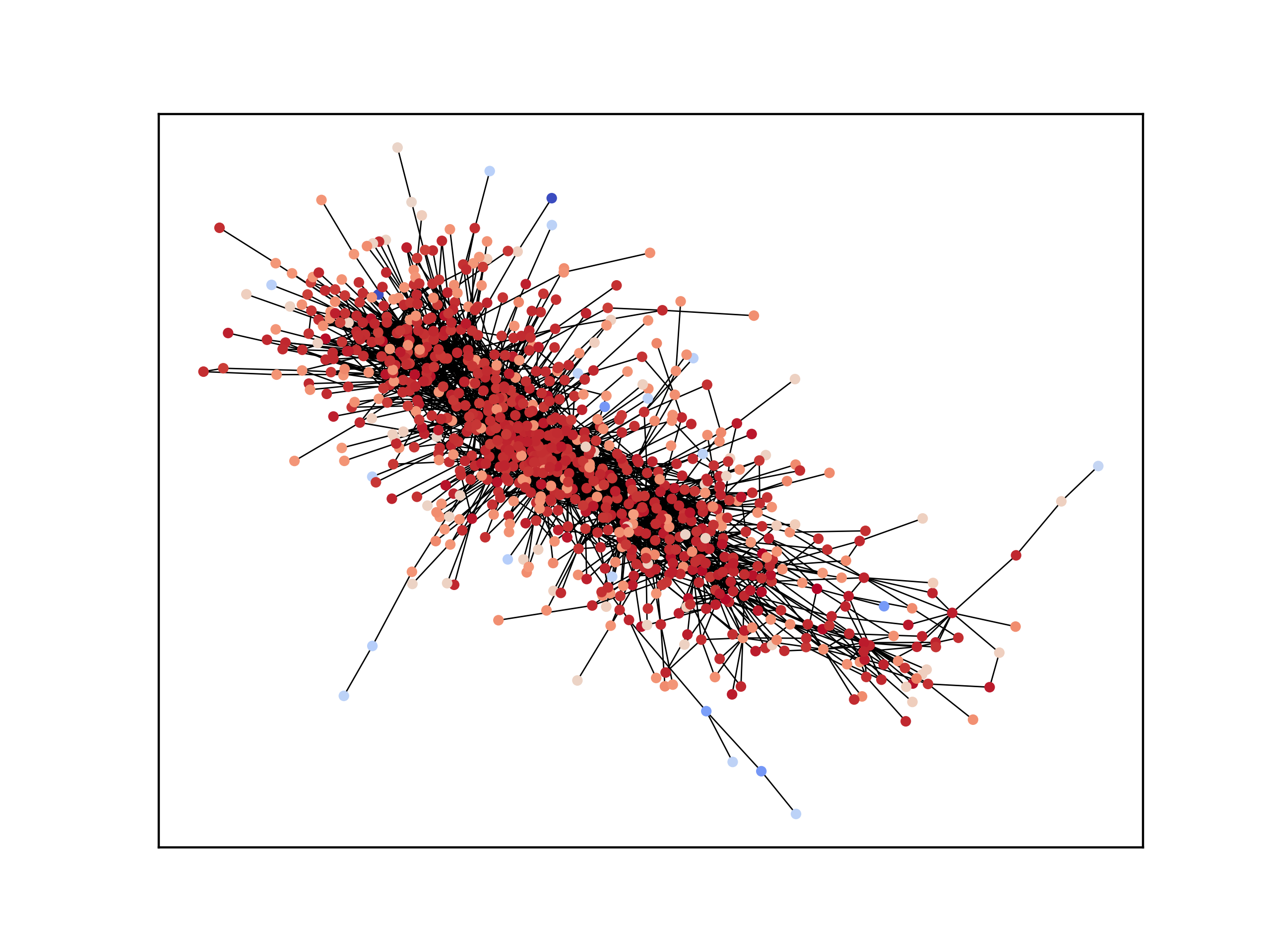}
    \label{fig:networks1}
  \end{subfigure}
  \begin{subfigure}{0.49\textwidth}
    \centering
    \includegraphics[width=\textwidth]{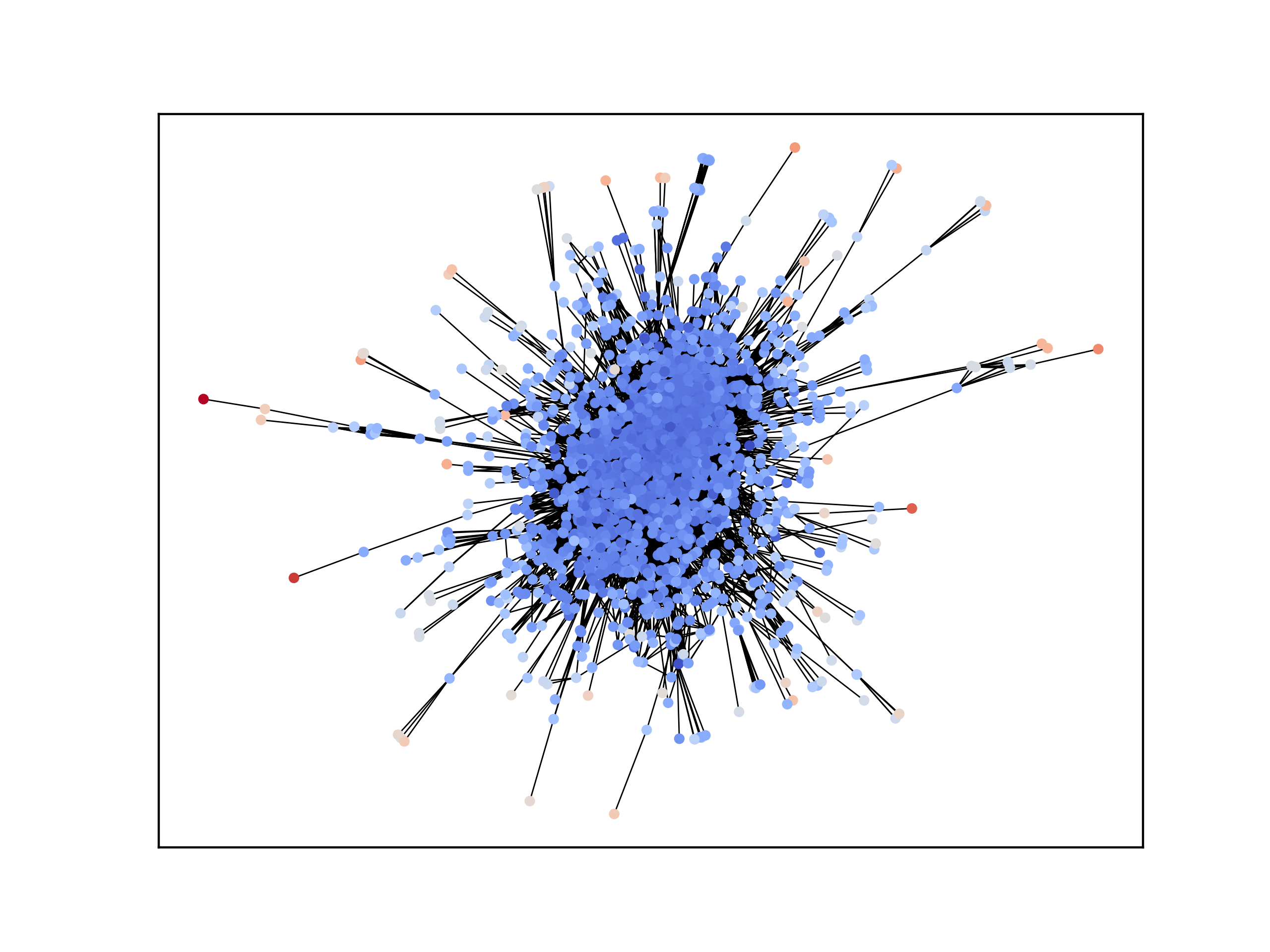}
    \label{fig:networks2}
  \end{subfigure}
  \caption{Visualization of Advogato (left) and Hamsterster (right) networks. The color represents the target value we get when spreading starts from a given node. Color on Advogato dataset represents the maximum number of infected nodes while on Hamsterster dataset time until the maximum number of infected nodes is reached is shown. Blue colors represent low values while red ones represent high ones. Since nodes with similar centrality values have similar characteristics, these nodes should be colored similarly.}
  \label{fig:networks}
\end{figure}

We used the following datasets for testing\footnote{Available at \url{https://github.com/smeznar/Epidemic-spreading-CN2020}.}: Hamsterster~\cite{soc-hamsterster}, Advogato~\cite{massa2009bowling}, Wikipedia Vote~\cite{leskovec2010signed}, FB Public Figures~\cite{rozemberczki2019gemsec}, and HEP-PH~\cite{leskovec2007hepph} taken from the Network Repository website~\cite{nr}. Some basic statistics of the networks we used can be seen in Table~\ref{tab:networks}. Two of the networks used during testing are visualized in Figure~\ref{fig:networks}. The network nodes in this figure are colored based on the values of the target variables. 

\begin{table}[h!]
\centering
\caption{Basic statistics of the networks used for testing.}
\resizebox{1\columnwidth}{!}{
\begin{tabular}{lrrcc}
\hline
Name & Nodes & Edges & Components & \makecell{Percentage of nodes \\in largest component}\\
\hline
Wikipedia Vote~\cite{leskovec2010signed}&   889&   2914&   1& 1\\
Hamsterster~\cite{soc-hamsterster}&   2426&   16630&  148 & 0.82\\
Advogato~\cite{massa2009bowling}&   6551&   43427&   1441 & 0.77\\
FB Public Figures~\cite{rozemberczki2019gemsec}&    11565&   67114&  1 & 1\\
HEP-PH~\cite{leskovec2007hepph}& 12008 & 118521 & 278 & 0.93\\
\hline
\end{tabular}}
\label{tab:networks}
\end{table}

We used the following approach to test the proposed method as well as the baselines mentioned in Section~\ref{sec:baselines}. We created the target data by simulating ten epidemics starting from each node of every dataset. We created each simulation using the SIR diffusion model from the NDlib~\cite{ndlib} Python library with parameters $\beta = 5\%$ and $\gamma = 0.5\%$. We then created the target variables by identifying and aggregating the maximum number of infected nodes and the time when this happens. We used these target variables to test the methods using five-fold cross-validation. We used XGBoost~\cite{xgboost} with default parameters as the regression model with proposed features based on the mentioned centralities, the random baseline, SNoRe~\cite{meznar2020snore}, SNoRe+centrality features, node2vec~\cite{grover2016node2vec}, and node2vec+centrality features baselines. Baselines GIN and GAT were trained for 200 epochs using the Adam optimizer~\cite{kingma2015Adam}. Since GIN and GAT are primarily used for node classification, we changed the output layer to a ReLU~\cite{nair2010relu} layer, so they perform regression.

\subsection{Results of models trained with simulation data}
\label{sec:results}
The results of the evaluation described in Section~\ref{sec:setting} are presented in Tables~\ref{tab:score-whole},~\ref{tab:time-whole},~\ref{tab:score-biggest}, and~\ref{tab:time-biggest}. Tables~\ref{tab:score-whole} and \ref{tab:time-whole} show the results on all the nodes from the network, while Tables~\ref{tab:score-biggest} and~\ref{tab:time-biggest} only on the nodes from the network's largest component. The results show that the learners significantly outperform the random baseline and the averaged simulation error, especially when predicting effects on networks with several components. Models CaBoost, node2vec+features, and SNoRe+features perform significantly better than  others and all use centrality-based features to train the XGBoost model. These best approaches achieve RMSE scores around 0.05, which corresponds to an error of around 5\% of nodes on average. 

The results for the prediction of the maximum number of infected nodes on the whole network are shown in Table~\ref{tab:score-whole}. The results show that SNoRe+features model has the lowest RMSE on most networks, but that this is mostly because of the centrality based features, since all the learners that use them give similar results. We also see that graph neural networks perform poorly, on most networks only beating the random baseline. This might be because we use features extracted from network and small amount of training data. It is also worth mentioning that the three best performing models perform notably better than the averaged simulation error and that the node embedding methods node2vec and SNoRe perform much worse when used without the centrality based features.

\begin{table*}[t!]
    \centering
    	\caption{Cross-validation results for maximum number of infected nodes on the whole network.}
    \resizebox{\columnwidth}{!}{
	\begin{tabular}{llllll}
\toprule
Dataset & Advogato & Hamsterster & HEP-PH & FB Public Figures & Wikipedia Vote \\
Learner           &                        &                        &                        &                        &                        \\
\midrule
CaBoost           &  0.0519 ($\pm$ 0.0045) &  \bfseries 0.0429 ($\pm$ 0.0116) &  0.0481 ($\pm$ 0.0017) &  0.0521 ($\pm$ 0.0007) &  0.0600 ($\pm$ 0.0020) \\
GAT               &  0.1748 ($\pm$ 0.0072) &  0.1534 ($\pm$ 0.0024) &  0.1761 ($\pm$ 0.0019) &  0.0594 ($\pm$ 0.0010) &  0.0608 ($\pm$ 0.0013) \\
GIN               &  0.0646 ($\pm$ 0.0238) &  0.0712 ($\pm$ 0.0597) &  0.1753 ($\pm$ 0.0806) &  0.0579 ($\pm$ 0.0015) &  0.2076 ($\pm$ 0.2531) \\
Random            &  0.3156 ($\pm$ 0.0024) &  0.2915 ($\pm$ 0.0029) &  0.2107 ($\pm$ 0.0012) &  0.0625 ($\pm$ 0.0003) &  0.0732 ($\pm$ 0.0039) \\
SNoRe             &  0.1743 ($\pm$ 0.0057) &  0.1591 ($\pm$ 0.0053) &  0.1611 ($\pm$ 0.0048) &  0.0600 ($\pm$ 0.0003) &  0.0667 ($\pm$ 0.0032) \\
SNoRe+features    &  \bfseries 0.0515 ($\pm$ 0.0044) &  0.0438 ($\pm$ 0.0114) &  \bfseries 0.0467 ($\pm$ 0.0018) &  \bfseries 0.0514 ($\pm$ 0.0004) &  0.0597 ($\pm$ 0.0009) \\
node2vec          &  0.0673 ($\pm$ 0.0054) &  0.0841 ($\pm$ 0.0143) &  0.0835 ($\pm$ 0.0031) &  0.0575 ($\pm$ 0.0005) &  0.0690 ($\pm$ 0.0021) \\
node2vec+features &  0.0574 ($\pm$ 0.0037) &  0.0431 ($\pm$ 0.0114) &  0.0494 ($\pm$ 0.0031) &  0.0515 ($\pm$ 0.0004) &  \bfseries 0.0590 ($\pm$ 0.0010) \\ \hline
Simulation error & 0.0644 & 0.0576 & 0.0796 & 0.0982 & 0.1064 \\
\bottomrule
\end{tabular}
}
	\label{tab:score-whole}
\end{table*}

Table~\ref{tab:time-whole} shows the performance results for prediction of time needed to reach the maximum number of infected nodes on the whole network. We see that SNoRe+features model performs the best overall. This is probably due to features that represent both the similarity between neighborhoods of nodes and their global characteristics. The results also show that GIN and GAT are not suitable for such a task since they often perform much worse than some other learners (especially GAT) and in some cases worse than the simulation error.

\begin{table*}[b!]
    \centering
    	\caption{Cross-validation results for time when maximum number of infected nodes is reached on the whole network.}
    \resizebox{\columnwidth}{!}{
	\begin{tabular}{llllll}
\toprule
Dataset & Advogato & Hamsterster & HEP-PH & FB Public Figures & Wikipedia Vote \\
Learner           &                        &                        &                        &                        &                        \\
\midrule
CaBoost           &  0.0571 ($\pm$ 0.0032) &  \bfseries 0.0540 ($\pm$ 0.0037) &  0.0459 ($\pm$ 0.0012) &  0.0448 ($\pm$ 0.0005) &  0.0647 ($\pm$ 0.0027) \\
GAT               &  0.1411 ($\pm$ 0.0021) &  0.1027 ($\pm$ 0.0012) &  0.0971 ($\pm$ 0.0006) &  0.0642 ($\pm$ 0.0008) &  0.0760 ($\pm$ 0.0021) \\
GIN               &  0.0782 ($\pm$ 0.0274) &  0.0766 ($\pm$ 0.0162) &  0.0717 ($\pm$ 0.0142) &  0.0497 ($\pm$ 0.0035) &  0.0701 ($\pm$ 0.0069) \\
Random            &  0.2073 ($\pm$ 0.0013) &  0.1209 ($\pm$ 0.0014) &  0.1095 ($\pm$ 0.0003) &  0.0817 ($\pm$ 0.0004) &  0.0992 ($\pm$ 0.0019) \\
SNoRe             &  0.1463 ($\pm$ 0.0033) &  0.1007 ($\pm$ 0.0022) &  0.0904 ($\pm$ 0.0006) &  0.0646 ($\pm$ 0.0011) &  0.0753 ($\pm$ 0.0073) \\
SNoRe+features    &  \bfseries 0.0557 ($\pm$ 0.0038) &  0.0545 ($\pm$ 0.0014) &  \bfseries 0.0451 ($\pm$ 0.0006) &  \bfseries 0.0434 ($\pm$ 0.0004) &  0.0641 ($\pm$ 0.0037) \\
node2vec          &  0.0758 ($\pm$ 0.0014) &  0.0824 ($\pm$ 0.0039) &  0.0634 ($\pm$ 0.0016) &  0.0590 ($\pm$ 0.0009) &  0.0845 ($\pm$ 0.0018) \\
node2vec+features &  0.0602 ($\pm$ 0.0032) &  0.0600 ($\pm$ 0.0035) &  0.0467 ($\pm$ 0.0015) &  0.0440 ($\pm$ 0.0004) &  \bfseries 0.0638 ($\pm$ 0.0016) \\ \hline
Simulation error & 0.0840 & 0.0906 & 0.0839 & 0.0847 & 0.1178 \\
\bottomrule
\end{tabular}
}
	\label{tab:time-whole}
\end{table*}

Similarly to Table~\ref{tab:score-whole}, Table~\ref{tab:score-biggest} shows the prediction scores for the maximum number of infected nodes on the largest component of the network. Results for networks Wikipedia vote and FB Public Figures are the same, since they have only one component. Contrary to scores on the whole network, scores on the biggest component show that node2vec+features performs the best overall. We also see that the random baseline performs much better on the single component than on the whole network. This is because the maximum number of infected nodes is usually smaller in smaller components, which makes the mean value of target data smaller and the variance higher. Because of high variance of target data, the random baseline predicts scores with higher error since the range of predictions is bigger.

\begin{table*}[t!]
    \centering
	\caption{Cross-validation results for maximum number of infected nodes on the biggest component of the network.}    \resizebox{\columnwidth}{!}{
	\begin{tabular}{llllll}
\toprule
Dataset & Advogato & Hamsterster & HEP-PH & FB Public Figures & Wikipedia Vote \\
Learner           &                        &                        &                        &                        &                        \\
\midrule
CaBoost           &  0.0556 ($\pm$ 0.0011) &  \bfseries 0.0437 ($\pm$ 0.0017) &  0.0496 ($\pm$ 0.0004) &  0.0521 ($\pm$ 0.0007) &  0.0600 ($\pm$ 0.0020) \\
GAT               &  0.0668 ($\pm$ 0.0088) &  0.0455 ($\pm$ 0.0014) &  0.0536 ($\pm$ 0.0014) &  0.0594 ($\pm$ 0.0010) &  0.0608 ($\pm$ 0.0013) \\
GIN               &  0.0651 ($\pm$ 0.0039) &  0.0566 ($\pm$ 0.0057) &  0.1147 ($\pm$ 0.0355) &  0.0579 ($\pm$ 0.0015) &  0.2076 ($\pm$ 0.2531) \\
Random            &  0.0622 ($\pm$ 0.0007) &  0.0513 ($\pm$ 0.0014) &  0.0556 ($\pm$ 0.0005) &  0.0625 ($\pm$ 0.0003) &  0.0732 ($\pm$ 0.0039) \\
SNoRe             &  0.0588 ($\pm$ 0.0005) &  0.0520 ($\pm$ 0.0014) &  0.0551 ($\pm$ 0.0003) &  0.0600 ($\pm$ 0.0003) &  0.0667 ($\pm$ 0.0032) \\
SNoRe+features    &  0.0552 ($\pm$ 0.0007) &  0.0447 ($\pm$ 0.0008) & \bfseries 0.0482 ($\pm$ 0.0004) &  \bfseries 0.0514 ($\pm$ 0.0004) &  0.0597 ($\pm$ 0.0009) \\
node2vec          &  0.0592 ($\pm$ 0.0008) &  0.0504 ($\pm$ 0.0012) &  0.0520 ($\pm$ 0.0002) &  0.0575 ($\pm$ 0.0005) &  0.0690 ($\pm$ 0.0021) \\
node2vec+features &  \bfseries 0.0548 ($\pm$ 0.0010) &  \bfseries 0.0437 ($\pm$ 0.0016) &  0.0489 ($\pm$ 0.0003) &  0.0515 ($\pm$ 0.0004) &  \bfseries 0.0590 ($\pm$ 0.0010) \\ \hline
Simulation error & 0.0975 & 0.0769 & 0.0883 & 0.0982 & 0.1064 \\
\bottomrule
\end{tabular}
}
	\label{tab:score-biggest}
\end{table*}

Table~\ref{tab:time-biggest} shows the prediction score of time needed to reach the maximum number of infected nodes on the biggest component of the network. As with the other results, CaBoost, node2vec+features, and SNoRe+features give the best performance on all dataset. Compared to results in Table~\ref{tab:time-whole}, we see that the difference between the random baseline and other learners is smaller and that the random baseline results are in some cases only around 50\% worse than the best performing learner. Interestingly, the random baseline gives better results overall than the averaged simulation error. This is probably because spreading is `highly' stochastic and simulations can end before spreading begins. In such case, the averaged simulation error increases significantly, while the random baseline is not affected much since the model is trained with already processed target data.\footnote{If we chose a random value $x\in[0, 1]$ as the prediction for the node, the result would be much worse.}

\begin{table*}[t!]
    \centering
	\caption{Cross-validation results for time when maximum number of infected nodes is reached on the biggest component of the network.}    \resizebox{\columnwidth}{!}{
	\begin{tabular}{llllll}
\toprule
Dataset & Advogato & Hamsterster & HEP-PH & FB Public Figures & Wikipedia Vote \\
Learner           &                        &                        &                        &                        &                        \\
\midrule
CaBoost           &  0.0529 ($\pm$ 0.0018) &  \bfseries 0.0442 ($\pm$ 0.0020) &  0.0436 ($\pm$ 0.0006) &  0.0448 ($\pm$ 0.0005) &  0.0647 ($\pm$ 0.0027) \\
GAT               &  0.0790 ($\pm$ 0.0063) &  0.0883 ($\pm$ 0.0031) &  0.0643 ($\pm$ 0.0008) &  0.0642 ($\pm$ 0.0008) &  0.0760 ($\pm$ 0.0021) \\
GIN               &  0.0614 ($\pm$ 0.0016) &  0.0536 ($\pm$ 0.0058) &  0.0538 ($\pm$ 0.0118) &  0.0497 ($\pm$ 0.0035) &  0.0701 ($\pm$ 0.0069) \\
Random            &  0.0855 ($\pm$ 0.0010) &  0.0907 ($\pm$ 0.0014) &  0.0845 ($\pm$ 0.0003) &  0.0817 ($\pm$ 0.0004) &  0.0992 ($\pm$ 0.0019) \\
SNoRe             &  0.0702 ($\pm$ 0.0011) &  0.0680 ($\pm$ 0.0026) &  0.0651 ($\pm$ 0.0005) &  0.0646 ($\pm$ 0.0011) &  0.0753 ($\pm$ 0.0073) \\
SNoRe+features    &  \bfseries 0.0517 ($\pm$ 0.0014) &  0.0454 ($\pm$ 0.0015) &  \bfseries 0.0425 ($\pm$ 0.0003) &  \bfseries 0.0434 ($\pm$ 0.0004) &  0.0641 ($\pm$ 0.0037) \\
node2vec          &  0.0708 ($\pm$ 0.0010) &  0.0702 ($\pm$ 0.0018) &  0.0557 ($\pm$ 0.0006) &  0.0590 ($\pm$ 0.0009) &  0.0845 ($\pm$ 0.0018) \\
node2vec+features &  0.0519 ($\pm$ 0.0017) &  0.0471 ($\pm$ 0.0030) &  0.0431 ($\pm$ 0.0002) &  0.0440 ($\pm$ 0.0004) &  \bfseries 0.0638 ($\pm$ 0.0016) \\ \hline
Simulation error & 0.0941 & 0.0789 & 0.0816 & 0.0847 & 0.1178 \\
\bottomrule
\end{tabular}
}
	\label{tab:time-biggest}
\end{table*}

We can see that predictions with the proposed learners on all datasets give better results than a single simulation. This shows that such models are useful because they can estimate the joint distribution of spreading across multiple simulations which is better than a random simulation run.

\subsection{Results of transfer learning}
\label{sec:transfer-results}
In this section, we show the results of transfer learning between the presented networks. The results are represented in the form of a heatmap where the values on the diagonal represent the baseline RMSE of five-fold cross-validation and the non-diagonal values the RMSE of the dataset in the column with the model trained on the dataset in the row. The error of non-diagonal cells is calculated on all nodes and divided by the baseline score and thus shows how much worse the RMSE we get from transfer learning is when compared with the RMSE we get with the five-fold cross-validation.

\begin{figure}[t]
  \centering
  \includegraphics[width=\linewidth]{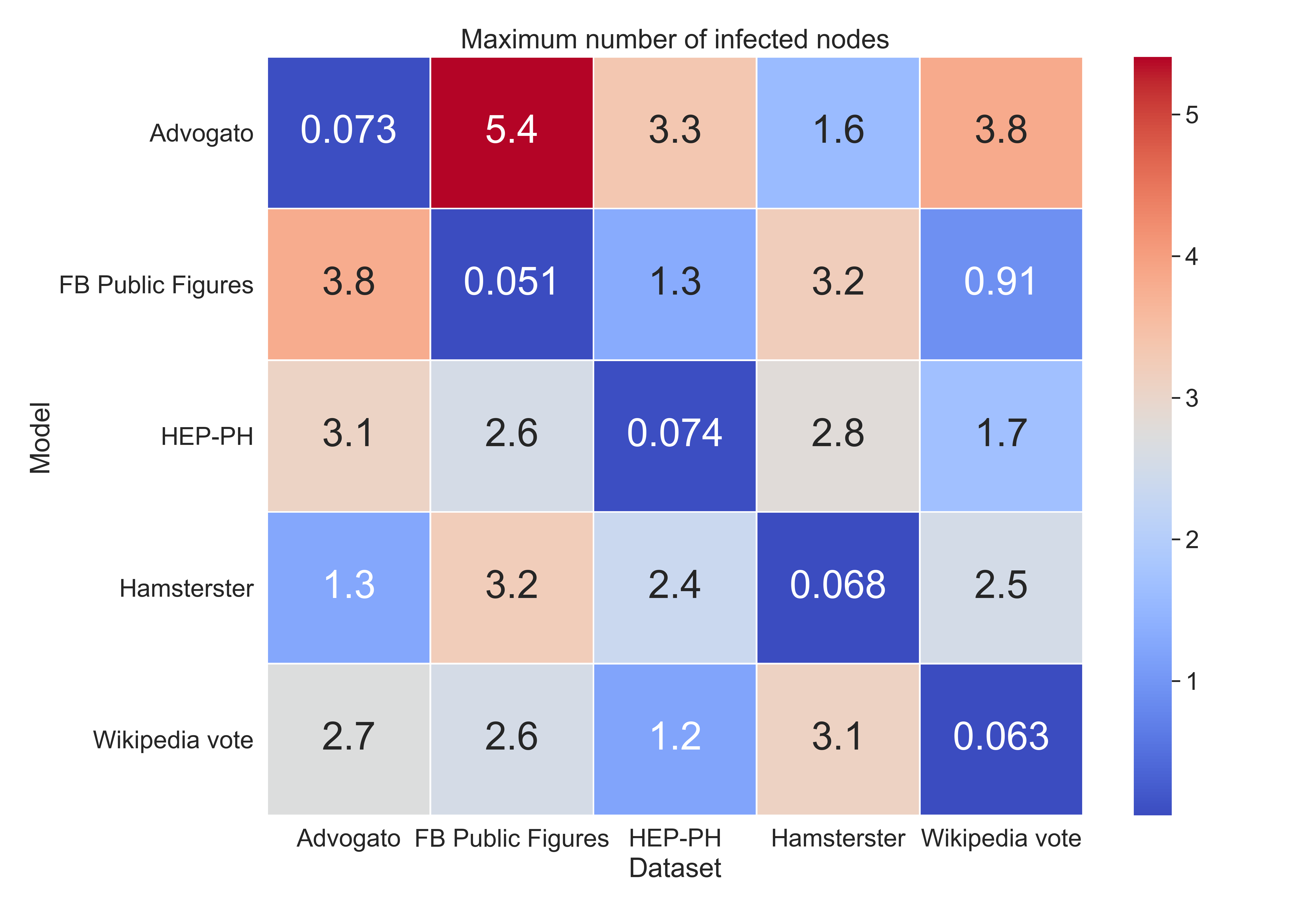}
  \caption{Heatmap with transfer learning results for predictions of maximum number of infected nodes.}
  \label{fig:transfer-score}
\end{figure}

The transfer learning results for the prediction of the maximum number of infected nodes can be seen on the heatmap in  Figure~\ref{fig:transfer-score}. We can see that most errors are 1--3 times higher than the baseline. The two major exceptions are the results of Advogato dataset with the FB Public Figures model and the result of Wikipedia Vote network with the FB Public Figures model. The 5.4 times higher RMSE on the FB Public Figure dataset is probably caused by the big difference between the number of components, since the large number of components lowers the highest number of infected nodes. It is interesting to see that the FB Public Figures model works better than the baseline for the Wikipedia network. This is probably because both networks have similar structure but Wikipedia vote has less nodes and thus less training data. These results show that transfer learning between two topologically similar networks is possible without additional data and can yield good results.

\begin{figure}[t]
  \centering
  \includegraphics[width=\linewidth]{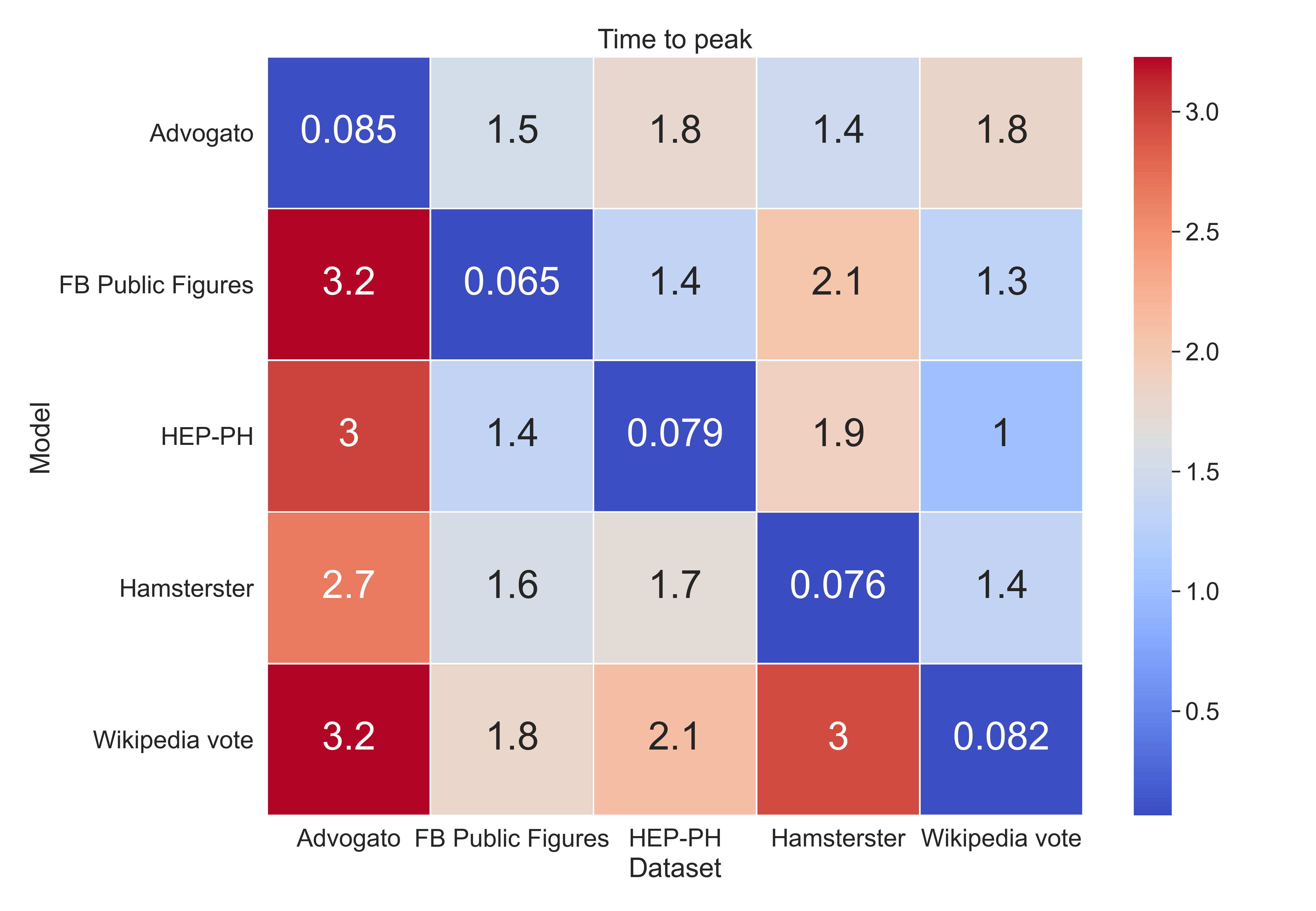}
  \caption{Heatmap with transfer learning results for predictions of time when maximum number of nodes were infected.}
  \label{fig:transfer-time}
\end{figure}

On the heatmap in Figure~\ref{fig:transfer-time} we see transfer learning results for prediction of the time needed to reach the maximum number of infected nodes. We can see that overall these results are better than those in Figure~\ref{fig:transfer-score} and that dataset Advogato performs much worse with other models. This is probably because Advogato has 1441 components while the other networks have significantly less. We can also see that FB Public Figures and Wikipedia vote datasets give good predictions (below 2 times worse) with all the models, especially in the case where the error is the same as with the baseline.

\begin{figure}[t!]
  \centering
  \includegraphics[width=\linewidth]{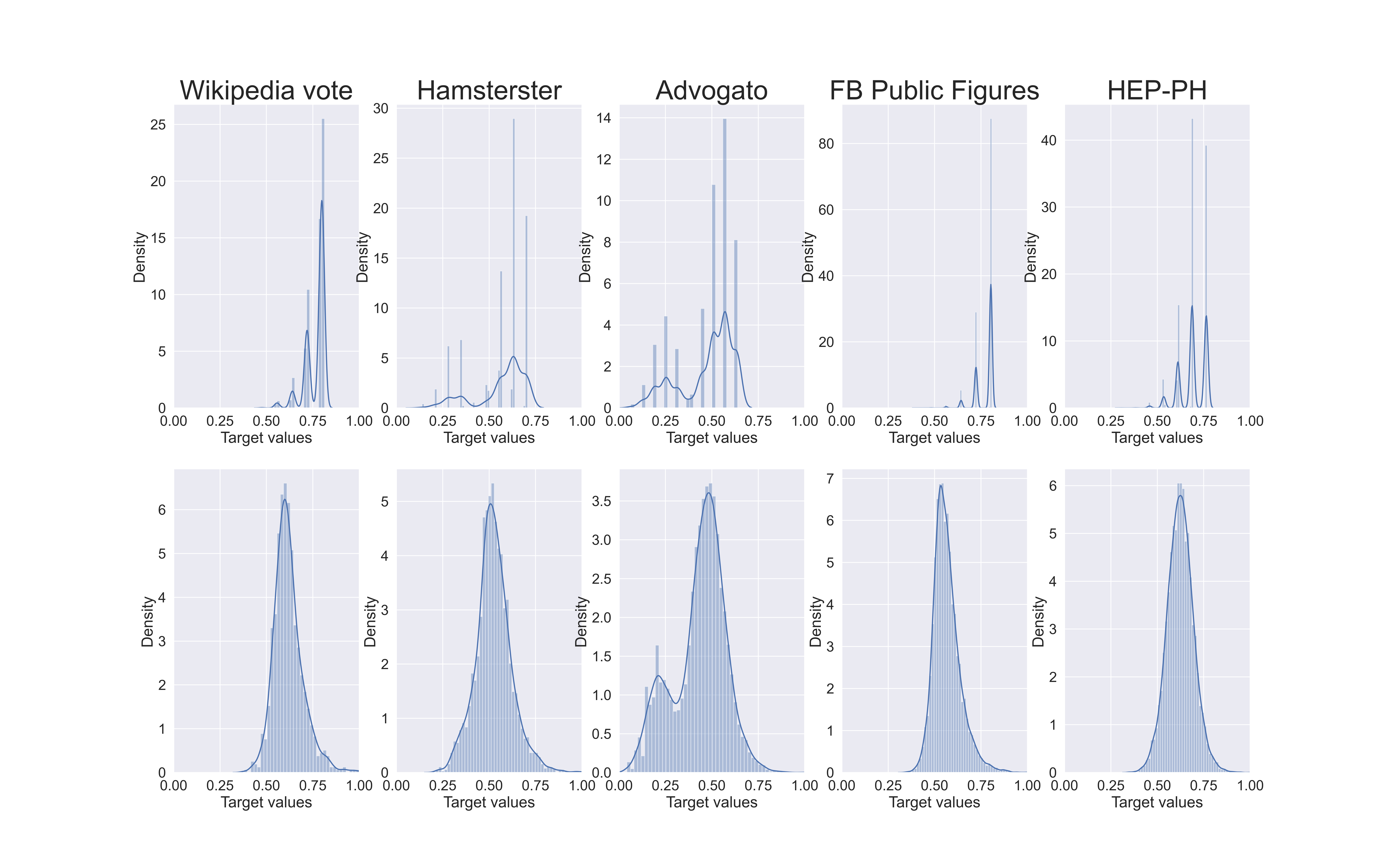}
  \caption{Distribution of target values for the maximum number of infected nodes and the time when this happens. The first row shows the target values for the number of infected nodes while the second one the time when this happens. The $x$-axis represents the value of the target variable. On the other hand the values of the $y$ axis represent the density at some value.}
  \label{fig:distribution}
\end{figure}

The results of transfer learning can be better explained with the distribution plot of target values shown in Figure~\ref{fig:distribution}. The first row shows the distribution of the maximum number of infected nodes. We can see that the distributions of FB Public Figures and Wikipedia vote are very similar. This reflects the results, where Wikipedia vote network performs better with the FB Public Figures model than with the five-fold cross-validation. We also see that the Advogato and FB Public Figures networks have very different distributions. This matches the results since the transferred model performs very poorly.

The second row shows the distribution of time needed to reach the maximum number of infected nodes. As with the maximum number of infected nodes these distributions also show that the distribution is closely related to how well the model performs. We see that the distribution of target values on the Advogato network vastly differs from the distributions on other networks and that this reflects the results where transfer learning models have higher RMSE. Similarly, the distributions of datasets FB Public Figures, HEP-PH, and Wikipedia vote are similar and have transfer learning results that do not differ much from the five-fold cross-validation results. 

\subsection{Interpretation of a prediction}

\begin{figure}[h!]
  \centering
  \includegraphics[width=.9\linewidth]{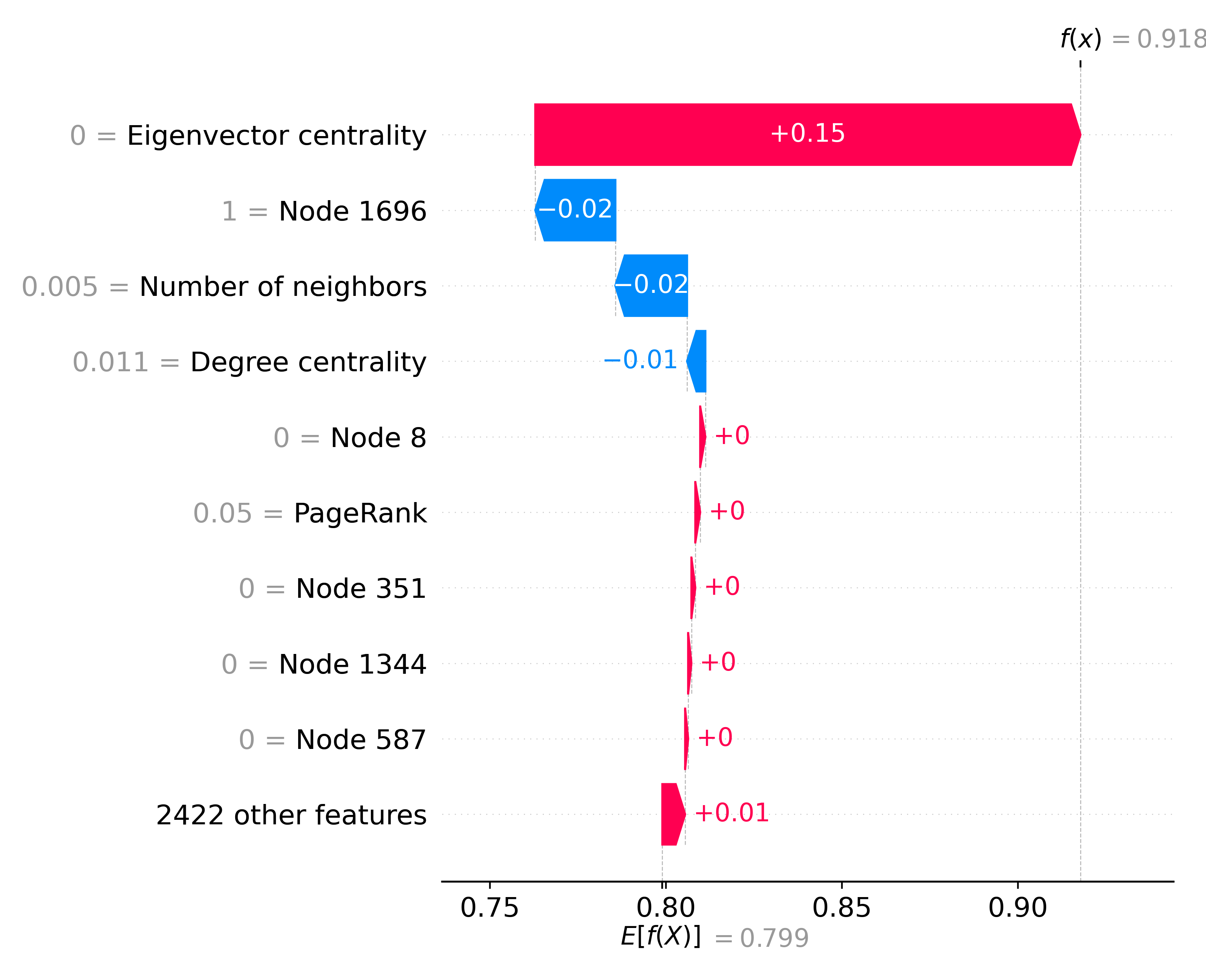}
  \caption{An example of a model explanation for an instance using the SNoRe+features model. Blue arrows indicate how much the prediction is lowered by some feature value, while the red ones indicate how much it is raised. Prediction starts at models expected value 0.799 and finishes at 0.918. Features and their values are shown on the left. The visualization shows for example that the prediction rose from 0.788 to 0.918 \emph{because of the low value of Eigenvector centrality}.}
  \label{fig:shap}
\end{figure}

We can explain predictions using model explanation approaches such as SHapley Additive exPlanations (SHAP)~\cite{lundberg2017shap,vstrumbelj2014explaining}. SHAP is a game-theoretic approach for explaining classification and regression models. The algorithm perturbs subsets of input features to take into account the interactions and redundancies between them. The explanation model can then be visualized, showing how the feature values of an instance impact a prediction.

An example of such explanation is shown in Figure~\ref{fig:shap} using the model SNoRe+features. We can see that the prediction is impacted mostly by the eigenvector centrality, node 1696, number of second neighbors, and the degree centrality. We can also see that a very small value of eigenvector centrality raises the prediction value and that the low values of the number of second neighbors and the Degree centrality lower it. This is expected because the low value of eigenvector centrality usually tells that the node is not that `important' and is in a neighborhood with many nodes. Similarly it is expected that the low value of degree centrality and low number of second neighbors lower the prediction because having less nodes gives a smaller chance of infection. Lastly, the high similarity between neighborhoods of node 1696 and the instance we try to predict lowers the prediction.

\section{Discussion and conclusions}
\label{sec:discussion-conclusion}
In this paper, we showcase that machine learning methods can be used for fast estimation of epidemic spreading effect from a given node. We show that by reformulating the task as node regression, we can obtain realistic estimations much faster than by performing computationally expensive simulations, even though such simulations are initially used to fine-tune the machine learning models. Further, employment of predictive modeling instead on relying on a single simulation also shows promising results. We also demonstrate that transfer learning can be used to predict spreading effects between networks with similar characteristics without big performance loss.

We show that while graph neural networks outperform the random baseline and can give us great results, centrality scores and embedding feature representation methods coupled with XGBoost mostly outperform them. We also see that machine learning models might overall give a more accurate representation of an epidemic than data gathered from a small number of simulations. This makes the machine learning approach faster and more reliable, while also giving an interpretation of why a node was predicted as it was. Further, this paper demonstrates the complementarity between the accepted simulation-based spreading modeling and  fast machine learning based screening in data-scarce regimes.

A crucial part of our work shows that transfer of knowledge between networks is possible. This implies that our features capture characteristics that are crucial and transferable between different networks. Since we derive features for models from centralities that are explainable, machine learning models can be used to study which characteristics of the networks play a crucial role in epidemic spreading and how they affect it.

An obvious limitation of the proposed task is that the spreading is probabilistic and even the best classifiers might make significant errors. Similarly, when observing prediction results of the maximum number of infected nodes one must be careful since we predict an average outcome from some node and not the true maximum. This gives us the ability to predict which nodes are the most `dangerous' as the patient zero. When trying to predict an outcome of an epidemic that has already spread, one must adjust data accordingly and get rid of simulations where epidemics have not spread.

In further work, we plan to research different centralities and algorithms to better describe network structure and achieve more accurate results. The proposed approach lowers the amount of simulations needed to create good approximations, but the approach might still not be scalable to larger networks. In the future, we would like to develop methods to further reduce the number of simulations needed, making the solution more scalable. Another area of our interest is the ability to solve such tasks by using unsupervised algorithms. Finally, as the current work is focused on the node-level aspects, we believe that similar ideas could be adopted to model higher order structures and their spreading potential, including convex skeletons and communities.

\section*{Acknowledgments}
The work of the first author was funded by the Slovenian Research Agency through a young researcher grant (B\v{S}).
The work of other authors was supported by the Slovenian Research Agency (ARRS) core research programs P2-0103 and P6-0411, 
and research projects J7-7303, L7-8269, and N2-0078 (financed under the ERC Complementary Scheme).

\bibliography{ms.bib}

\begin{thebibliography}{10}

\bibitem{xgboost}
Tianqi Chen and Carlos Guestrin.
\newblock {XGBoost: A Scalable Tree Boosting System}.
\newblock In {\em Proceedings of the 22nd ACM SIGKDD International Conference
  on Knowledge Discovery and Data Mining}, KDD '16, page 785–794, New York,
  NY, USA, 2016. Association for Computing Machinery.

\bibitem{dong2018studies}
Suyalatu Dong, Feng-Hua Fan, and Yong-Chang Huang.
\newblock Studies on the population dynamics of a rumor-spreading model in
  online social networks.
\newblock {\em Physica A: Statistical Mechanics and its Applications},
  492:10--20, 2018.

\bibitem{Fey2019ptg}
Matthias Fey and Jan~E. Lenssen.
\newblock {Fast Graph Representation Learning with PyTorch Geometric}.
\newblock In {\em ICLR Workshop on Representation Learning on Graphs and
  Manifolds}, 2019.

\bibitem{Granovetter1979threshold}
Mark Granovetter.
\newblock {Threshold Models of Collective Behavior}.
\newblock {\em American Journal of Sociology}, 83(6):1420--1443, 1978.

\bibitem{grover2016node2vec}
Aditya Grover and Jure Leskovec.
\newblock node2vec: Scalable feature learning for networks.
\newblock In Balaji Krishnapuram, Mohak Shah, Alexander~J. Smola, Charu~C.
  Aggarwal, Dou Shen, and Rajeev Rastogi, editors, {\em Proceedings of the 22nd
  {ACM} {SIGKDD} International Conference on Knowledge Discovery and Data
  Mining, San Francisco, CA, USA, August 13-17, 2016}, pages 855--864. {ACM},
  2016.

\bibitem{Guille2013diffusion}
Adrien Guille, Hakim Hacid, Cecile Favre, and Djamel~A. Zighed.
\newblock {Information Diffusion in Online Social Networks: A Survey}.
\newblock {\em SIGMOD Rec.}, 42(2):17–28, 2013.

\bibitem{soc-hamsterster}
Hamsterster.
\newblock Hamsterster social network.
\newblock http://www.hamsterster.com.

\bibitem{hu2019strategies}
Weihua Hu, Bowen Liu, Joseph Gomes, Marinka Zitnik, Percy Liang, Vijay Pande,
  and Jure Leskovec.
\newblock Strategies for pre-training graph neural networks.
\newblock {\em arXiv preprint arXiv:1905.12265}, 2019.

\bibitem{kacem2017small}
Ahmed Kacem, Christine Lallemand, Nathalie Giraud, Maxime Mense, Matthieu
  De~Gennaro, Yannick Pizzo, Jean-Claude Loraud, Pascal Boulet, and Bernard
  Porterie.
\newblock A small-world network model for the simulation of fire spread onboard
  naval vessels.
\newblock {\em Fire Safety Journal}, 91:441--450, 2017.

\bibitem{Kampffmeyer_2019_CVPR}
Michael Kampffmeyer, Yinbo Chen, Xiaodan Liang, Hao Wang, Yujia Zhang, and
  Eric~P. Xing.
\newblock Rethinking knowledge graph propagation for zero-shot learning.
\newblock In {\em {IEEE} Conference on Computer Vision and Pattern Recognition,
  {CVPR} 2019, Long Beach, CA, USA, June 16-20, 2019}, pages 11487--11496.
  Computer Vision Foundation / {IEEE}, 2019.

\bibitem{Kempe2003cascades}
David Kempe, Jon Kleinberg, and \'{E}va Tardos.
\newblock {Maximizing the Spread of Influence through a Social Network}.
\newblock In {\em Proceedings of the Ninth ACM SIGKDD International Conference
  on Knowledge Discovery and Data Mining}, KDD '03, page 137–146, New York,
  NY, USA, 2003. Association for Computing Machinery.

\bibitem{Kermack1927Epidemics}
William~Ogilvy Kermack, A.~G. McKendrick, and Gilbert~Thomas Walker.
\newblock A contribution to the mathematical theory of epidemics.
\newblock {\em Proceedings of the Royal Society of London. Series A, Containing
  Papers of a Mathematical and Physical Character}, 115(772):700--721, 1927.

\bibitem{kesarev2018parallel}
Sergey Kesarev, Oksana Severiukhina, and Klavdiya Bochenina.
\newblock Parallel simulation of community-wide information spreading in online
  social networks.
\newblock In {\em Russian Supercomputing Days}, pages 136--148. Springer, 2018.

\bibitem{kingma2015Adam}
Diederik~P. Kingma and Jimmy Ba.
\newblock Adam: {A} method for stochastic optimization.
\newblock In Yoshua Bengio and Yann LeCun, editors, {\em 3rd International
  Conference on Learning Representations, {ICLR} 2015, San Diego, CA, USA, May
  7-9, 2015, Conference Track Proceedings}, 2015.

\bibitem{kipf2016semi}
Thomas~N. Kipf and Max Welling.
\newblock Semi-supervised classification with graph convolutional networks.
\newblock In {\em 5th International Conference on Learning Representations,
  {ICLR} 2017, Toulon, France, April 24-26, 2017, Conference Track
  Proceedings}. OpenReview.net, 2017.

\bibitem{lee2021transfer}
Chee-Kong Lee, Chengqiang Lu, Yue Yu, Qiming Sun, Chang-Yu Hsieh, Shengyu
  Zhang, Qi~Liu, and Liang Shi.
\newblock Transfer learning with graph neural networks for optoelectronic
  properties of conjugated oligomers.
\newblock {\em The Journal of Chemical Physics}, 154(2):024906, 2021.

\bibitem{leskovec2010signed}
Jure Leskovec, Daniel~P. Huttenlocher, and Jon~M. Kleinberg.
\newblock Signed networks in social media.
\newblock In Elizabeth~D. Mynatt, Don Schoner, Geraldine Fitzpatrick, Scott~E.
  Hudson, W.~Keith Edwards, and Tom Rodden, editors, {\em Proceedings of the
  28th International Conference on Human Factors in Computing Systems, {CHI}
  2010, Atlanta, Georgia, USA, April 10-15, 2010}, pages 1361--1370. {ACM},
  2010.

\bibitem{leskovec2007hepph}
Jure Leskovec, Jon Kleinberg, and Christos Faloutsos.
\newblock Graph evolution: Densification and shrinking diameters.
\newblock {\em ACM Trans. Knowl. Discov. Data}, 1(1):2–es, 2007.

\bibitem{li2017survey}
Mei Li, Xiang Wang, Kai Gao, and Shanshan Zhang.
\newblock A survey on information diffusion in online social networks: Models
  and methods.
\newblock {\em Information}, 8(4):118, 2017.

\bibitem{lundberg2017shap}
Scott~M. Lundberg and Su{-}In Lee.
\newblock A unified approach to interpreting model predictions.
\newblock In Isabelle Guyon, Ulrike von Luxburg, Samy Bengio, Hanna~M. Wallach,
  Rob Fergus, S.~V.~N. Vishwanathan, and Roman Garnett, editors, {\em Advances
  in Neural Information Processing Systems 30: Annual Conference on Neural
  Information Processing Systems 2017, December 4-9, 2017, Long Beach, CA,
  {USA}}, pages 4765--4774, 2017.

\bibitem{Mallick2020TransferLW}
Tanwi Mallick, Prasanna Balaprakash, E.~Rask, and J.~MacFarlane.
\newblock Transfer learning with graph neural networks for short-term highway
  traffic forecasting.
\newblock {\em ArXiv}, abs/2004.08038, 2020.

\bibitem{massa2009bowling}
Paolo Massa, Martino Salvetti, and Danilo Tomasoni.
\newblock Bowling alone and trust decline in social network sites.
\newblock In {\em Dependable, Autonomic and Secure Computing, 2009. DASC'09.
  Eighth IEEE International Conference on}, pages 658--663. IEEE, 2009.

\bibitem{meznar2020snore}
S.~{Me\v{z}nar}, N.~{Lavra\v{c}}, and B.~{\v{S}krlj}.
\newblock Snore: Scalable unsupervised learning of symbolic node
  representations.
\newblock {\em IEEE Access}, 8:212568--212588, 2020.

\bibitem{meznar2020spreading}
Sebastian Me{\v{z}}nar, Nada Lavra{\v{c}}, and Bla{\v{z}} {\v{S}}krlj.
\newblock Prediction of the effects of epidemic spreading with graph neural
  networks.
\newblock In Rosa~M. Benito, Chantal Cherifi, Hocine Cherifi, Esteban Moro,
  Luis~Mateus Rocha, and Marta Sales-Pardo, editors, {\em Complex Networks {\&}
  Their Applications IX}, pages 420--431, Cham, 2021. Springer International
  Publishing.

\bibitem{nair2010relu}
Vinod Nair and Geoffrey~E. Hinton.
\newblock Rectified linear units improve restricted boltzmann machines.
\newblock In Johannes F{\"{u}}rnkranz and Thorsten Joachims, editors, {\em
  Proceedings of the 27th International Conference on Machine Learning
  (ICML-10), June 21-24, 2010, Haifa, Israel}, pages 807--814. Omnipress, 2010.

\bibitem{nowzari2016analysis}
Cameron Nowzari, Victor~M Preciado, and George~J Pappas.
\newblock {Analysis and control of epidemics: A survey of spreading processes
  on complex networks}.
\newblock {\em IEEE Control Systems Magazine}, 36(1):26--46, 2016.

\bibitem{Page1999ThePC}
L.~Page, S.~Brin, R.~Motwani, and T.~Winograd.
\newblock {The PageRank Citation Ranking: Bringing Order to the Web.}
\newblock In {\em WWW 1999}, 1999.

\bibitem{Perozzi2014deepwalk}
Bryan Perozzi, Rami Al{-}Rfou, and Steven Skiena.
\newblock Deepwalk: online learning of social representations.
\newblock In Sofus~A. Macskassy, Claudia Perlich, Jure Leskovec, Wei Wang, and
  Rayid Ghani, editors, {\em The 20th {ACM} {SIGKDD} International Conference
  on Knowledge Discovery and Data Mining, {KDD} '14, New York, NY, {USA} -
  August 24 - 27, 2014}, pages 701--710. {ACM}, 2014.

\bibitem{qiu2020gcc}
Jiezhong Qiu, Qibin Chen, Yuxiao Dong, Jing Zhang, Hongxia Yang, Ming Ding,
  Kuansan Wang, and Jie Tang.
\newblock Gcc: Graph contrastive coding for graph neural network pre-training.
\newblock In {\em Proceedings of the 26th ACM SIGKDD International Conference
  on Knowledge Discovery \& Data Mining}, pages 1150--1160, 2020.

\bibitem{rodrigues2019network}
Francisco~Aparecido Rodrigues.
\newblock {Network Centrality: An Introduction}.
\newblock {\em A Mathematical Modeling Approach from Nonlinear Dynamics to
  Complex Systems}, page 177, 2019.

\bibitem{ndlib}
Giulio Rossetti, Letizia Milli, Salvatore Rinzivillo, Alina S{\^\i}rbu, Dino
  Pedreschi, and Fosca Giannotti.
\newblock {NDlib}: a python library to model and analyze diffusion processes
  over complex networks.
\newblock {\em International Journal of Data Science and Analytics},
  5(1):61--79, 2018.

\bibitem{nr}
Ryan~A. Rossi and Nesreen~K. Ahmed.
\newblock The network data repository with interactive graph analytics and
  visualization.
\newblock In Blai Bonet and Sven Koenig, editors, {\em Proceedings of the
  Twenty-Ninth {AAAI} Conference on Artificial Intelligence, January 25-30,
  2015, Austin, Texas, {USA}}, pages 4292--4293. {AAAI} Press, 2015.

\bibitem{rozemberczki2019gemsec}
Benedek Rozemberczki, Ryan Davies, Rik Sarkar, and Charles Sutton.
\newblock {GEMSEC:} graph embedding with self clustering.
\newblock In Francesca Spezzano, Wei Chen, and Xiaokui Xiao, editors, {\em
  {ASONAM} '19: International Conference on Advances in Social Networks
  Analysis and Mining, Vancouver, British Columbia, Canada, 27-30 August,
  2019}, pages 65--72. {ACM}, 2019.

\bibitem{sge}
Bla{\v{z}} {\v{S}}krlj, Nada Lavra{\v{c}}, and Jan Kralj.
\newblock {{Symbolic Graph Embedding Using Frequent Pattern Mining}}.
\newblock In Petra Kralj~Novak, Tomislav {\v{S}}muc, and Sa{\v{s}}o
  D{\v{z}}eroski, editors, {\em Discovery Science}, pages 261--275, Cham, 2019.
  Springer International Publishing.

\bibitem{vstrumbelj2014explaining}
Erik {\v{S}}trumbelj and Igor Kononenko.
\newblock Explaining prediction models and individual predictions with feature
  contributions.
\newblock {\em Knowledge and information systems}, 41(3):647--665, 2014.

\bibitem{velickovic2018graph}
Petar Velickovic, Guillem Cucurull, Arantxa Casanova, Adriana Romero, Pietro
  Li{\`{o}}, and Yoshua Bengio.
\newblock Graph attention networks.
\newblock In {\em 6th International Conference on Learning Representations,
  {ICLR} 2018, Vancouver, BC, Canada, April 30 - May 3, 2018, Conference Track
  Proceedings}. OpenReview.net, 2018.

\bibitem{wang2019zerosurvey}
Wei Wang, Vincent~W. Zheng, Han Yu, and Chunyan Miao.
\newblock A survey of zero-shot learning: Settings, methods, and applications.
\newblock {\em ACM Trans. Intell. Syst. Technol.}, 10(2), 2019.

\bibitem{wu2020comprehensive}
Zonghan Wu, Shirui Pan, Fengwen Chen, Guodong Long, Chengqi Zhang, and S~Yu
  Philip.
\newblock A comprehensive survey on graph neural networks.
\newblock {\em IEEE Transactions on Neural Networks and Learning Systems},
  2020.

\bibitem{zhu2002learning}
Zhu Xiaojin and Ghahramani Zoubin.
\newblock Learning from labeled and unlabeled data with label propagation.
\newblock {\em Tech. Rep., Technical Report CMU-CALD-02--107, Carnegie Mellon
  University}, 2002.

\bibitem{xu2018powerful}
Keyulu Xu, Weihua Hu, Jure Leskovec, and Stefanie Jegelka.
\newblock How powerful are graph neural networks?
\newblock In {\em 7th International Conference on Learning Representations,
  {ICLR} 2019, New Orleans, LA, USA, May 6-9, 2019}. OpenReview.net, 2019.

\bibitem{zhuang2020comprehensive}
Fuzhen Zhuang, Zhiyuan Qi, Keyu Duan, Dongbo Xi, Yongchun Zhu, Hengshu Zhu, Hui
  Xiong, and Qing He.
\newblock A comprehensive survey on transfer learning.
\newblock {\em Proceedings of the IEEE}, 109(1):43--76, 2020.

\end{thebibliography}
\bibliographystyle{plain}

\end{document}